 \documentclass[12pt]{article}
 \usepackage{amssymb}
 \usepackage{graphicx}
 \usepackage{psfrag}
 \usepackage{amsfonts}
 \usepackage{epsfig}
 \usepackage{rotate}
 \usepackage{latexsym}

 \newcommand{\sef}{\sin^2 \theta_{eff}^{lept}}
 \newcommand{\ini}{\begin{equation}}
 \newcommand{\fin}{\end{equation}}

 \newcommand{\es}{s_{eff}^2}

 \newcommand{\dr}{\Delta r}

 \newcommand{\dah}{\Delta \alpha_h^{(5)}}
 
 \newcommand{\asmz}{\alpha_s \left( M_Z \right)}
 
 \newcommand{\bmath}{\begin{displaymath}}
 \newcommand{\emath}{\end{displaymath}}
 \newcommand{\bite}{\begin{itemize}}
 \newcommand{\eite}{\end{itemize}}
 \renewcommand\thefootnote{\fnsymbol{footnote}}

\begin{document}

 \hyphenation{re-nor-ma-li-za-tion}

 \begin{flushright}
 NYU-TH/04-06-28\\
 Freiburg-THEP 04/11\\
 hep-ph/0406334
 \end{flushright}

 \vspace{0.5cm}
 \begin{center}
 \boldmath{\Large \bf Bounds on the Higgs Boson Mass from $M_W$\footnote{Talk given by G.~Ossola at the Incontri sulla Fisica delle 
Alte Energie, XVI Ciclo di Incontri, Torino, Italy, April 14-16, 2004.}
 }\unboldmath\\
 \vspace{0.2cm}
 \vspace{0.5cm}
 {\large
 A.~Ferroglia$^{a}$,
 G.~Ossola$^{b}$,
 and A.~Sirlin$^{b}$}
  

 \vspace{0.5cm}
 $^a${\it Fakult\"at f\"ur Physik,  Universit\"at Freiburg,\\ D-79104 Freiburg, Germany} \\
 \vspace{0.5cm}
 $^b${\it Department of Physics, New York University,\\
 4 Washington Place, New York, NY 10003, USA}
 \end{center}
 \bigskip
 \bigskip
 \bigskip
 \renewcommand\thefootnote{\arabic{footnote}}
 \addtocounter{footnote}{-1}

 \begin{center}
 \bf Abstract
 \end{center}
 { \small Recent experimental and theoretical progress in the $M_H$ estimates from $M_W$
   and $\sef$ is reviewed, with particular emphasis on the role played by $M_t$.
   Assuming that the SM is correct and taking into account the lower bound on $M_H$
   from direct searches, we derive restrictive bounds on $M_W$ and $M_t$.
   We also discuss a representative ``benchmark'' scenario for the possible future
   evolution of these parameters. Amusingly, this benchmark scenario suggested some
   time ago a value for $M_t$ that turned out to be in very close agreement with its
   most recent experimental determination.
 }

 \newpage

 \section{Introduction}

 The Standard Model of Particle Physics (SM) has been very successful in
 describing the interactions among elementary particles 
 involving the fundamental forces, with the exception of gravity.
 Although at present no experimental data are in sharp 
 contradiction with the predictions of this theory, 
 many questions remain open. 
 In particular, one important component of the theory, the Higgs boson, 
 responsible via the Higgs mechanism for the
 generation of the masses of all fundamental 
 particles, has yet to be discovered. 

 The precise electroweak experiments performed at the colliders  (LEP, SLD, C0,
 CDF) can be used to check the validity of the  Standard Model and, within its
 framework, to get important informations about its parameters. The high
 accuracy of the measurements makes their interpretation sensitive to quantities
 that appear in the electroweak  corrections, such as the mass of the top quark
 $M_t$ and the mass of the Higgs boson $M_H$. 
 The leading dependence of the electroweak corrections on $M_t$ and $M_H$ is
 quadratic and logarithmic, respectively; as a consequence, it is more difficult
 to put sharp constraints on $M_H$ that on $M_t$.
 In the case of the top quark, the analysis of the $Z^0$-resonance observables
 led to a rather precise prediction of $M_t$, before its experimental discovery.
 This indirect prediction turned out to be in good agreement with the value of
 $M_t$ measured at Fermilab in 1995, giving strong evidence for the presence of
 electroweak corrections. Although it is still interesting and important to
 derive indirect predictions for the masses of already discovered particles, such
 as $M_W$ and $M_t$, at present the main focus of the indirect analysis is on the
 determination of the allowed ranges for the great missing piece, the mass of the
 Higgs boson.
 %

 The EWWG (LEP Electroweak Working Group) performs a wide set of
 $\chi^2$-minimization fits to compare all the available experimental data with
 the theoretical predictions of the SM
 \cite{lepewwg}. This procedure, known as a global
 analysis, provides important information, tests the validity of the theory, and
 attempts to find deviations that may signal the presence of new physics beyond
 the SM
 \cite{glo_ifae}.
  On the other hand, it has been argued that, aside from the global
 analyses, it is also important to confront the theory separately with the
 precise observables that are most sensitive to $M_H$, such as $M_W$ and $\sef$
 \cite{cx3,chano2,paolo2003}.
 In fact, it is in principle possible that striking discrepancies between crucial
 observables and important information may be blurred in the global analysis 
 \cite{marciano_sirlinfest}.

 In the following, we will first discuss the bounds on $M_H$
  that can be obtained solely from the experimental value of $M_W$,
 giving particular attention to the role played in this analysis by $M_t$.
  In this context we will also review the main sources of 
 theoretical uncertainty.
 We will then comment briefly on the bounds that can be obtained using the 
 experimental value of $\sef$ and the unresolved issues involving the 
 current measurements of this important parameter. 
 We will finally present a different
 analysis in which, assuming that the SM is correct and taking into account the 
 lower limit on $M_H$ from direct searches, we derive restrictive 
 bounds on $M_W$, $M_t$.
 Analogous bounds for $\sef$ have been obtained \cite{bounds}. 

 \boldmath
 \section{Bounds on the Higgs Boson Mass from $M_W$}
 \unboldmath

 There are several factors that single out the $M_W$ determination as
 particularly important:
 \bite
 \item The LEP2 and Tevatron experimental measurements of
 $M_W$ are in excellent agreement;
 \item  As we will show in this Section, it places
 sharp restrictions on $M_H$;
 \item The relevant electroweak correction $\dr$ \cite{dr1} 
 has been now fully evaluated at the two-loop level \cite{m4}.
 \eite 
 In other words, in the $M_W$ case, we compare
 very precise theoretical results with highly consistent experimental 
 data.
 This is particularly important given the extreme sensitivity of these tests.
  Since the leading dependence of the theoretical formula
 for $M_W$ is logarithmic in $M_H$,  small shifts in the experimental values
 lead to large changes in $M_H$. As we mentioned before, 
 this is one of the reasons that complicates the indirect 
 determination of $M_H$ (in contrast, for example, to the analogous 
 determination of $M_t$).

 The recent experimental value for $M_W$, derived from the combination of
 the LEP2 and Tevatron results, is 
 \ini
 (M_W)_{exp} = 80.426 \pm 0.034\, \mbox{GeV}\, .
 \fin

 We will perform the fits using the theoretical  formula 
 introduced in  Ref.~\cite{sim}, 
 in the effective scheme of renormalization (EFF) \cite{eff}.
 The results  will be checked using also the 
 more complete formula by Awramik {\emph{ et al}}. \cite{cz}
 that takes into account the complete two-loop contributions
 to $\dr$.

 In order to stress the importance of $M_t$,
  we first perform the analysis using the Winter 2003 value
 of this parameter, i.e. $M_t = 174.3 \pm 5.1$ GeV, together with 
 $\dah = 0.02761 \pm 0.00036$ and $\asmz = 0.118 \pm 0.002$.
 Using the formula of Ref.~\cite{sim}, we obtain 
 \ini \label{6nog1}
 M_H = 45_{-36}^{+69}\, \mbox{GeV}\, ; \quad  M_H^{95} = 184\, \mbox{GeV}\, ,
 \label{ee2}
 \fin
 while employing the expressions  of Ref.~\cite{cz}, we have
 \ini \label{6nog2}
 M_H = 36_{-33}^{+65}\, \mbox{GeV}\, ; \quad  M_H^{95} =  168\, \mbox{GeV}\, .
 \label{ee3}
 \fin

 These results seem to suggest a very light Higgs boson.
  In fact, the central values in the predictions 
 of Eqs.~(\ref{6nog1}) and~(\ref{6nog2}) are well below
 the  95\% C.L. lower bound $(M_H)_{L.B.} = 114.4 \, \mbox{GeV}$
  from direct searches.
 On the other hand,  the 95\% C.L. upper bound $M_H^{95}$ in Eqs.~(\ref{ee2},
 \ref{ee3}) are still well above $(M_H)_{L.B.}$.
 Thus, solely from this analysis 
 we cannot conclude that there is clear evidence for inconsistencies and,
 consequently, need for ``new physics''. 
 It is interesting to compare the situation 
 depicted in~(\ref{6nog1}) and~(\ref{6nog2})
 with the corresponding analysis performed with the data available in 
 Winter 2002:
 using $(M_W)_{exp} = 80.451 \pm 0.033$ GeV (the experimental value for $M_W$ 
 in Winter 2002), we obtain in the EFF scheme \cite{sim}
 \ini \label{4e7}
 M_H = 23_{-23}^{+49}\, \mbox{GeV}\,;\quad  M_H^{95} = 122 \, \mbox{GeV}\, ,
 \fin
 a very worrisome low value!
 It is noteworthy to observe how a change in the central value of $(M_W)_{exp}$ of less then 
 1 $\sigma$, without changes in its error bar, has significantly improved 
 the consistency with the theory. 

 Before commenting on the effect of $M_t$, we briefly review the 
 sources of uncertainty in the theoretical calculation of $M_W$. There are two 
 types of theoretical errors,  one due to the truncation of the perturbative
 expansion (\emph{truncation error}) and one due to the uncertainties
 in the input parameters employed in  the calculation 
 (\emph{parametric error}).
 Including the effect of QCD corrections, the truncation error in the calculation
 of $M_W$ has been estimated to be $\approx 7 \, \mbox{MeV}$  in Ref.~\cite{sim}
 and $\approx 4 \, \mbox{MeV}$ in Ref.~\cite{cz}. 
 A parametric error of approximately the same size can 
 be obtained by shifting $\dah$ by about 1~$\sigma$ or $M_t$ by only $1\
 \mbox{GeV}$.
 We can therefore conclude that the main source of theoretical error is 
 still related to the uncertainty in the measurement 
 of $M_t$! A summary of the theoretical errors in the  calculations 
 of $M_W$ and $\sef$ is given in Table~\ref{tab1}.

 \begin{table}[ht]
     \caption{Parametric error in the calculation of $\es$ and  
       $M_W$ to be compared with a truncation error of $6 \times 10^{-5}$
       in $\es$ and  7 MeV (4 MeV) in  $M_W$.}
     \label{tab1}
     \begin{center}
     \begin{tabular}{ccccc}
       \hline
       \bf Parameter & 1$\sigma$ shift & $\Delta\es$  & $\Delta M_W$ \\ 
       \hline
       $\dah$ & 0.00036 & $1.3 \times 10^{-4}$ & 6 MeV \\ 
       $M_t$ & 4.3 GeV &  $1.4 \times 10^{-4}$ & 26 MeV \\
       \hline
     \end{tabular}
   \end{center}
 \end{table}

\begin{figure}[hp]
 \centering \psfrag{m}{ {\footnotesize $M_H$}}
 \psfrag{s}{{\footnotesize $M_W$ }}
 \psfig{figure=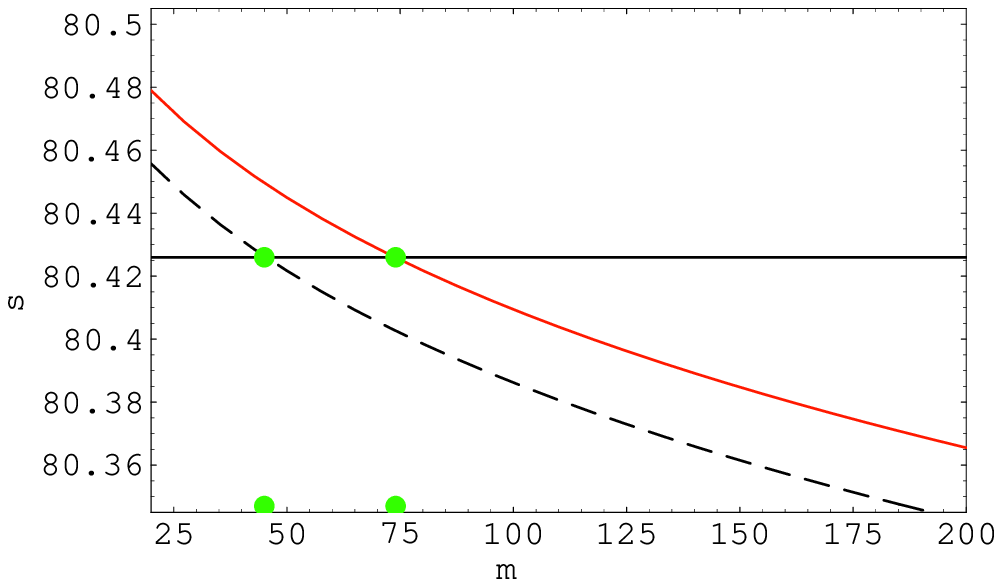,width=4truein,height=2.5truein} 
 \caption{The theoretical prediction for $M_W$ obtained with the new value of
 $M_t$ (full red line) is approximately 20 MeV higher then the 
 one obtained with the
 old $M_t$ value (dashed black line), 
 leading to an increase in the indirect 
 determination of $M_H$ (green dots).}\label{fig_mw}
 \end{figure}

 After a glimpse at Table~\ref{tab1}, 
 it is not surprising that the new experimental value   
 $M_t = 178.0 \pm 4.3$ GeV, has a big effect on the
 theoretical prediction of $M_W$ and therefore on the estimate of $M_H$.
 Repeating the analysis of Eqs.~(\ref{6nog1}) and~(\ref{6nog2}) with the 
 new value for $M_t$, we obtain
 \ini \label{6nog3}
 M_H = 74_{-47}^{+83}\, \mbox{GeV}\, ; \quad  M_H^{95} = 238 \, \mbox{GeV}
 \fin
 and
 \ini \label{6nog4}
 M_H = 62_{-43}^{+78}\, \mbox{GeV}\, ; \quad  M_H^{95} =  216\, \mbox{GeV}\, ,
 \fin 
 respectively, a significantly less restrictive range for $M_H$. 
The effect of the
change in $M_t$ is depicted in Fig.~\ref{fig_mw}, employing the theoretical
expression of Ref.~\cite{sim} in the effective scheme of renormalization.

 It will be very important to see in which direction the experimental values 
 of $M_W$ and $M_t$  evolve in future, more accurate, 
 experiments.
 Some plausible future scenarios are described in Section~\ref{bench}.

 \boldmath
 \section{The $\sef$ Anomaly}
 \unboldmath

 The analysis based on $\sef$ is more complicated.  Averaging all the different
 measurements, one obtains  $(\sef)_{exp} = 0.23150 \pm 0.00016$ \cite{lepewwg}
 with a  $\chi^2/\mbox{Dof} = 10.5/5$ corresponding to a confidence level of
 $6.2$\%. The averages derived from the leptonic and hadronic observables 
 are 
 $(\sef)_{(l)} = 0.23113 \pm 0.00021$ and
  $(\sef)_{(h)} = 0.23214 \pm 0.00027$, respectively, a difference of
 almost 3 $\sigma$! Thus, at present, the results obtained from the leptonic and 
 hadronic sectors are not in good agreement.

 On the theoretical side, a complete two-loop
 calculation for the prediction of $\es$ is not yet available.
 Our analysis employs the theoretical formula of 
 Ref.~\cite{sim} in the effective scheme of renormalization.
 In turn, this expression is based on the calculation reported in Ref.~\cite{eff}
 and includes two-loop effects enhanced by powers of $M_t^2/M_W^2$, as well as
 QCD corrections. 

 Using the world average $(\sef)_{exp} = 0.23150 \pm 0.00016$, together with the
 ``new'' value  $M_t = 178.0 \pm 4.3$ GeV, we find  \ini M_H = 159_{-61}^{+92}\,
 \mbox{GeV}\, ; \quad  M_H^{95} = 332\, \mbox{GeV}\, . \fin This result is in
 good agreement with the SM, in the sense that it suggests a light Higgs  boson
 with a mass above the region excluded by direct searches. However one should
 not forget  that the poor agreement between the leptonic and hadronic
 determinations of  $\sef$ is hidden in the average.\\ If we simply use the data
 from hadronic asymmetries, we obtain instead \ini M_H = 491_{-210}^{+342}\,
 \mbox{GeV}\, ; \quad  M_H^{95} = 1150\, \mbox{GeV}\, , \fin a significantly
 heavier Higgs boson. Finally using $(\sef)_{(l)} = 0.23113 \pm 0.00021$ from
 leptonic asymmetries, we find
 \ini
 M_H = 76_{-35}^{+58}\, \mbox{GeV}\, ; \quad  M_H^{95} = 190\, \mbox{GeV}\, ,
 \fin
 a result that is in very good agreement with that reported in Eq.~(\ref{6nog3})
 on the basis of the $M_W$ input. 

\begin{figure}[hp]
 \begin{center}
 \psfrag{m}{ {\small $M_H$ [GeV]}}
 \psfrag{s}{{\footnotesize $\sef$ }}
 \psfig{figure=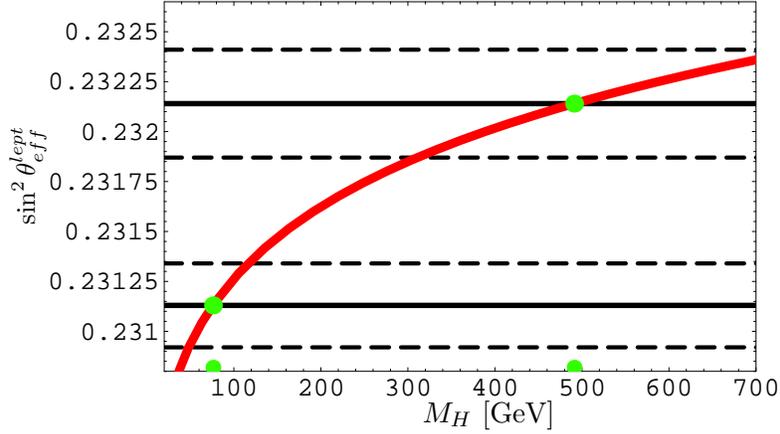,width=4truein,height=2.5truein}
 \caption{Comparison of the theoretical prediction for $\es$  with the experimental 
 averages derived from the leptonic observables ($\es = 0.23113 \pm 0.00021$) 
 and hadronic observables ($\es = 0.23214 \pm 0.00027$). The two values 
 lead to very different determinations of $M_H$ (green dots).}  \label{fig_seff}
 \end{center}
 \end{figure}

 We see that hadronic asymmetries  prefer a medium-heavy Higgs boson, while
 leptonic asymmetries, together with $M_W$, suggest
  a light particle. The situation is
 depicted in Fig.~\ref{fig_seff}.

 If we want to explain this contradiction, there are two 
 possibilities\footnote{This is known  in the literature as the ``Chanowitz
 argument''~\cite{chano1}, and has been  used in the past to stress the necessity
 of ``new physics'' beyond the SM. In the light of the new value for the top
 quark mass, of course, this argument is losing some of its original strength.}:
 \bite 
 \item The differences are due to statistical fluctuations, maybe
 enhanced   by unknown systematics.  
 \item The hadronic data is not described
 correctly by the SM.  
 \eite  
 The second possibility requires the introduction of some ``new physics'' in
 the  hadronic sector.  If the fluctuations were to settle on the side of the
 leptonic averages, the first possibility may also require new physics, since
 the predicted $M_H$ is smaller than $(M_H)_{L.B.}$. This will certainly be the
 case if in future experiments the central values remain as they are and the
 error bars are significantly reduced.

 As a final application, we combine the analysis based on $M_W$ and $\sef$,
 using the most recent $M_t$ value. If the world average 
 $\sef = 0.23150 \pm 0.00016$ is employed, we obtain 
 \ini
 M_H =138^{+80}_{-51}\, \mbox{GeV}\, ; \quad  M_H^{95} = 280\, \mbox{GeV}  \, .
 \fin
 Instead, if we restrict ourselves to the value 
 $(\sef)_{(l)} = 0.23113 \pm 0.00021$ derived from the leptonic observables, we
 have
 \ini
 M_H = 75_{-33}^{+54}\, \mbox{GeV}\, ; \quad  M_H^{95} = 177\, \mbox{GeV}  \, .
 \fin

 \boldmath 
 \section{Bounds on $M_W$ and $M_t$. A Benchmark Scenario}
 \unboldmath \label{bench}

 Assuming that the Standard Model is correct and taking into account
 the lower bound on $M_H$ from direct searches, we discuss
 bounds on $M_W$ and $M_t$ at various confidence levels.
 This permits to identify theoretically favored ranges for these important
 parameters in the Standard Model framework, regardless of other observables.
 This section is based on the work of Ref.~\cite{bounds}, in which 
  a similar analysis based on $\es$ and $M_t$ was also performed.

\begin{figure}[ht] \begin{center}
 \psfrag{m}{ {\footnotesize $M_t$}}
 \psfrag{s}{{\footnotesize $M_W$ }}
 \psfig{figure=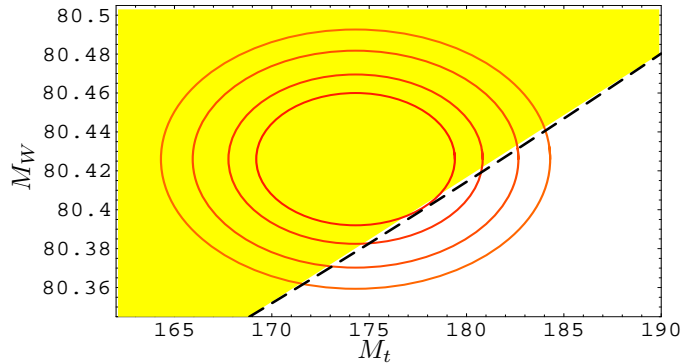,width=3.5truein,height=2truein} 
 \caption{Theoretical SM curve 
 $M_W(M_H,M_t)$ for $M_H = 114.4$ (dashed line) together with
 68\%, 80\%, 90\%, 95\% C.L. contours
 derived from $(M_W)_{exp}$ and $(M_t)_{exp}$ (Winter 2003).} \label{ell1}
 \end{center}
 \end{figure}

 Let us first consider the  data in Winter 2003 and
 $\dah = 0.02761 \pm 0.00036$. Fig.~\ref{ell1} shows the theoretical SM curve 
 $M_W(M_H,M_t)$ for $M_H = 114.4$ (dashed line) evaluated with the simple formulae
 of Ref.~\cite{sim} in the effective scheme of renormalization \cite{eff}, 
 as well as the 68\%, 80\%, 90\%, 95\% C.L. contours
 derived from the experimental values $(M_W)_{exp} = 80.426
 \pm 0.034 \, \mbox{GeV}$, $(M_t)_{exp} = 174.3 \pm 5.1 \,
 \mbox{GeV}$ (Winter 2003). 

 To simplify the analysis we take the restriction $M_H \ge 114.4\, \mbox{GeV}$
 to be a sharp cutoff rather than a $95\%$
 C.L. bound. At a given C.L. the allowed region lies within the corresponding
 ellipse and below the $M_H = 114.4\, \mbox{GeV}$ SM theoretical
 curve (dashed line), which we call the boundary curve (B.C.).
 It turns out that, to a good approximation, 
 the maximum and minimum $M_W$ and
 $M_t$ values in a given allowed region are determined by the
 intersections of the B.C. with the associated ellipse.
 The allowed $M_W$ and $M_t$ ranges determined by
 such intersections are shown in Table~\ref{tabell1} for the 80\%,
 90\%, 95\% C.L. domains. As $M_H$ increases beyond $114.4\,
 \mbox{GeV}$, the allowed ranges decrease in size.
 Clearly they are considerably more restrictive than the corresponding intervals
 derived from current experimental errors.

 \begin{table}[hp] 
 \caption{Allowed ranges for $M_W$ and $M_t$. 
 Results obtained in the EFF scheme with 
 $\dah = 0.02761$ and $(M_t)_{exp} = 174.3 \pm 5.1 \,
 \mbox{GeV}$ (Winter 2003). } \label{tabell1} 
 \begin{center}
 \begin{tabular}{ccc} 
 \hline
 EFF / $\dah = 0.02761$ & range $M_W$ [GeV]    &  range $M_t$ [GeV]    \\
 \hline
 80\% C.L. & 80.401 $\pm$ 0.018 & 177.9 $\pm$ 2.9 \\
 90\% C.L. & 80.401 $\pm$ 0.030 & 177.9 $\pm$ 4.8 \\
 95\% C.L. & 80.401 $\pm$ 0.040 & 177.9 $\pm$ 6.3 \\
 \hline
 \end{tabular} \end{center} 
 \end{table}
 The mid-points $(80.401, 177.9) \mbox{GeV}$ of the allowed regions in
 Table~\ref{tabell1} are independent of the C.L. and are shifted from the
 experimental central values by less then 1 $\sigma$. This  makes them
 particularly attractive representative points, which we identify as benchmarks
 to illustrate the possible future evolution of these fundamental parameters.

 This analysis was first performed in  Winter of 2003.
 It is amusing to observe how the benchmark point for $M_t$ 
  turned out to be very close
 to its new experimental central value. 

\begin{figure}[ht] 
 \begin{center}
 \psfrag{m}{ {\footnotesize $M_t$}}
 \psfrag{s}{{\footnotesize $M_W$ }}
 \psfig{figure=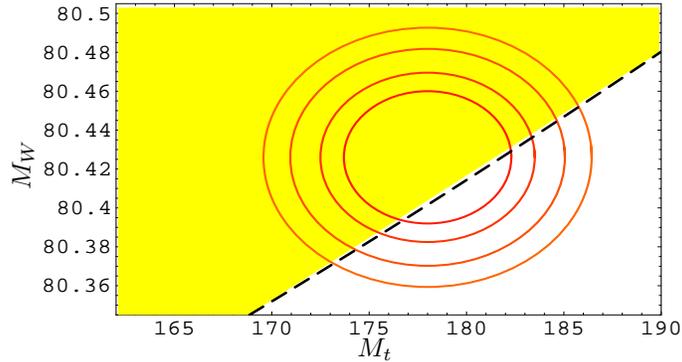,width=3.5truein,height=2truein}
 \caption{Same as Fig.~\ref{ell1}, using the new experimental data (April 2004).} \label{ell2}
 \end{center} 
 \end{figure}

 As a last application, we repeat the same analysis using the new 
 value for $M_t$ (Fig.~\ref{ell2} and Table~\ref{tabell2}).

 \begin{table}[ht] 
 \caption{Allowed ranges for $M_W$ and $M_t$. 
 Results obtained in the EFF scheme with 
 $\dah = 0.02761$ and $(M_t)_{exp} = 178.0 \pm 4.3 \,
 \mbox{GeV}$ (April 2004). }  \label{tabell2}
 \begin{center}
 \begin{tabular}{ccc}
 \hline
 EFF / $\dah = 0.02761$ & range $M_W$ [GeV]    &  range $M_t$ [GeV]    \\
 \hline
 68\% C.L. & 80.412 $\pm$ 0.018 & 179.5 $\pm$ 2.8 \\
 80\% C.L. & 80.412 $\pm$ 0.025 & 179.5 $\pm$ 3.9 \\
 90\% C.L. & 80.412 $\pm$ 0.034 & 179.5 $\pm$ 5.2 \\
 95\% C.L. & 80.412 $\pm$ 0.041 & 179.5 $\pm$ 6.3 \\
 \hline
 \end{tabular} \end{center}
 \end{table}
 Thus, the benchmark scenario currently prefers a further increase of 
 $\approx 1.5 \, \mbox{GeV}$ in $M_t$ and a value for $M_W$ that coincides with the
 most precise present measurement, namely the LEP2 central value $M_W = 80.412 \,
 \mbox{GeV}$.

 \section{Conclusions}

 At present, there is no compelling evidence of new physics beyond the SM.
 The new experimental value of $M_t$ has significantly improved the agreement
 between the SM and the experimental data and, in particular, the estimate of
 $M_H$ obtained via $M_W$. However, the central value of this predictions is still
 smaller than $(M_H)_{L.B.}$ by $\approx 40-50 \, \mbox{GeV}$. We emphasize that the
 $M_H$ analysis is very sensitive to $M_W$ and $M_t$. Thus, it will be very
 important to see how their values evolve in the future, as the experimental
 accuracy increases. 
In this connection, assuming that the SM is correct, we have
 derived bounds on $M_W$ and $M_t$ that are considerably more
 restrictive than the corresponding intervals derived from current experimental
 errors. We have also discussed a representative ``benchmark'' scenario for the
 possible future evolution of these parameters. Amusingly, this analysis
 suggested a value for $M_t$ that turned out to be in very close agreement with
 its most recent experimental determination.

 \section*{Acknowledgments}
 G.O. would like to thank R.~Chierici for interesting communications.
 The work of A.S. was supported in part by NSF Grant PHY-0245068.


 \end{document}